\documentclass[fleqn,12pt,twoside]{article}
\usepackage[headings]{espcrc1}

\usepackage{graphicx}

\newcommand{\be}{\begin{eqnarray}}
\newcommand{\ee}{\end{eqnarray}}
\def\lsim{\mathrel{\rlap{\lower3pt\hbox{\hskip1pt$\sim$}}
     \raise1pt\hbox{$<$}}} 
\def\gsim{\mathrel{\rlap{\lower3pt\hbox{\hskip1pt$\sim$}}
     \raise1pt\hbox{$>$}}} 

\title{ Polymer Chains and Baryons in a Strongly Coupled Quark-Gluon Plasma}

\author {Jinfeng Liao and
Edward V. Shuryak  
\address { Department of Physics and Astronomy\\
State University of New York,
     Stony Brook, NY 11794-3800}
}
\runtitle{Polymers and baryons in sQGP}
\runauthor{E.Shuryak}
\begin{document}

\date{\today}
\maketitle
\begin{abstract}
Recently there was a significant change of views on physical
properties
and underlying dynamics of Quark-Gluon Plasma at $T=170-350\, MeV$,
produced in heavy ion collisions at RHIC. Instead of being a gas of
$q,g$ quasiparticles,
 a near-perfect liquid is observed. Also, precisely in this
temperature interval, the interaction deduced from lattice studies
is strong enough to support multiple binary bound states. This
work is the first variational study of  {\em multibody} bound
states. We will consider: (i) ``polymer chains'' of the type $\bar
q g g ..g q$; (ii) baryons $(qqq)$; (iii) closed (3-)chains of
gluons $(ggg)$. We found that chains (i) form  in exactly the same
$T$ range as binary states, with the same binding {\em per bond}.
The binding and $T$-range for diquarks, baryons and closed
3-chains are also established. We point out that the presence of
chains, or possibly even a chain network, may
 drastically change the transport properties
of matter, such as charm diffusion or jet energy loss. We further
suggest that it seems to exist only for $T=(1-1.5)T_c$ and thus
there may be
 a ``latent period''  for charm/jet quenching in RHIC collisions,
while
matter
cools down to such $T$.
\end{abstract}

\section{Introduction}

 Quark-gluon plasma (QGP) is a high-temperature phase of QCD,
and the word ``plasma'' means
 the same sense as in electrodynamic plasma physics,
 the presence of free  (color) charges. QGP is different from hadronic
phase because charges
 are screened  by the medium \cite{Shu_QGP} rather than confined
in neutral objects. Lattice simulations have shown, that QGP
exists above the phase transition $T>T_c\approx 170 \, MeV$, and
is not only deconfined but also possesses the restored chiral
symmetry.

  Asymptotic freedom of non-Abelian gauge theories like QCD
  ensures
that for high enough temperature $T>>\Lambda_{QCD}$ this phase
becomes {\em weakly} coupled (wQGP), with most of interactions
characterized by small coupling $\alpha_s(p\sim T)<<1$. In this
domain wQGP is essentially a  near-ideal gas of its fundamental
constituents, quarks and gluons.

  QGP is experimentally studied via heavy ion collisions, at CERN SPS
and last years at BNL RHIC collider, reaching temperatures up to
about $T\approx 2T_c$. Success of hydrodynamical description
\cite{hydro} of observed collective flows has indicated, that all
dissipative lengths are very short and thus the produced matter
cannot be a weakly coupled gas but rather a very good liquid
\cite{Shu_liquid}. Recent studies of charm transport \cite{MT} and
preliminary RHIC data indicate, that charm diffusion constant is
also much smaller than pQCD predictions. This complements another
well known puzzle, of unexpectedly strong jet quenching at RHIC,
with its so far unexplained  angular dependence.

 In order to explain all of these features, a
 radically new picture of QGP at such temperatures
 is being developed, referred to as {\em strongly coupled} QGP, or sQGP.
 It has been in particular pointed
out in \cite{SZ_rethinking}
 that
the  interaction is strong enough to preserve the meson-like bound
states up to about $T=2T_c\sim 300\, MeV$, (the temperature range
corresponding to QGP at RHIC), although in a strongly modified
form. It was then pointed out in \cite{SZ_bound} that also
multiple binary {\em colored} bound states should exist in the
same $T$ domain. Since QGP is a deconfined phase, there is nothing
wrong with that, and the forces between say  singlet $\bar q q$
and octet $q g$ quasiparticle pairs are about the same. (For
potential-like forces the Casimir scaling gives 9/8 ratio, for
string-like ones the ratio is just 1.)
 Some of those states (charmonium) were observed
on the lattice \cite{charmonium} at $T$ up to about $2.5 T_c$,
while existence of most of these states, especially colored,
 still has to be checked.

  By this work we make the first step toward the understanding of the
{\em multibody} bound states. For definiteness, we will use
similar parameterized lattice-based
interactions as in \cite{SZ_bound}.

In the QCD vacuum, the potential
for two color charges is traditionally written as a sum of a Coulomb and
linear
potential, dominating small and large distances, respectively. At $T>T_c$,
 by definition of the deconfined phase, the effective string tension
vanishes and the potentials go to a constant at $r\rightarrow\infty$.
But that does not mean that string-like field configurations of color
field
disappear right at $T_c$:
as explained by Polyakov long ago \cite{Polyakov:1978vu},
the string tension which vanishes at $T=T_c$
should be understood as the free energy, $F=V-TS$, while the string
energy $V$ and its entropy (related to the number of configurations)
$S$ are finite but cancelling each other. This picture of the
deconfinement suggests by itself that the ``mixed phase'', at $T=T_c$,
contains a lot of very long strings. It is natural to think then, that
strong (although finite-range) interaction between the charges at $T>T_c$
is also related with strings.

\begin{figure}
\begin{minipage}{8.cm}
\vskip -.3cm
\includegraphics[width=8cm]{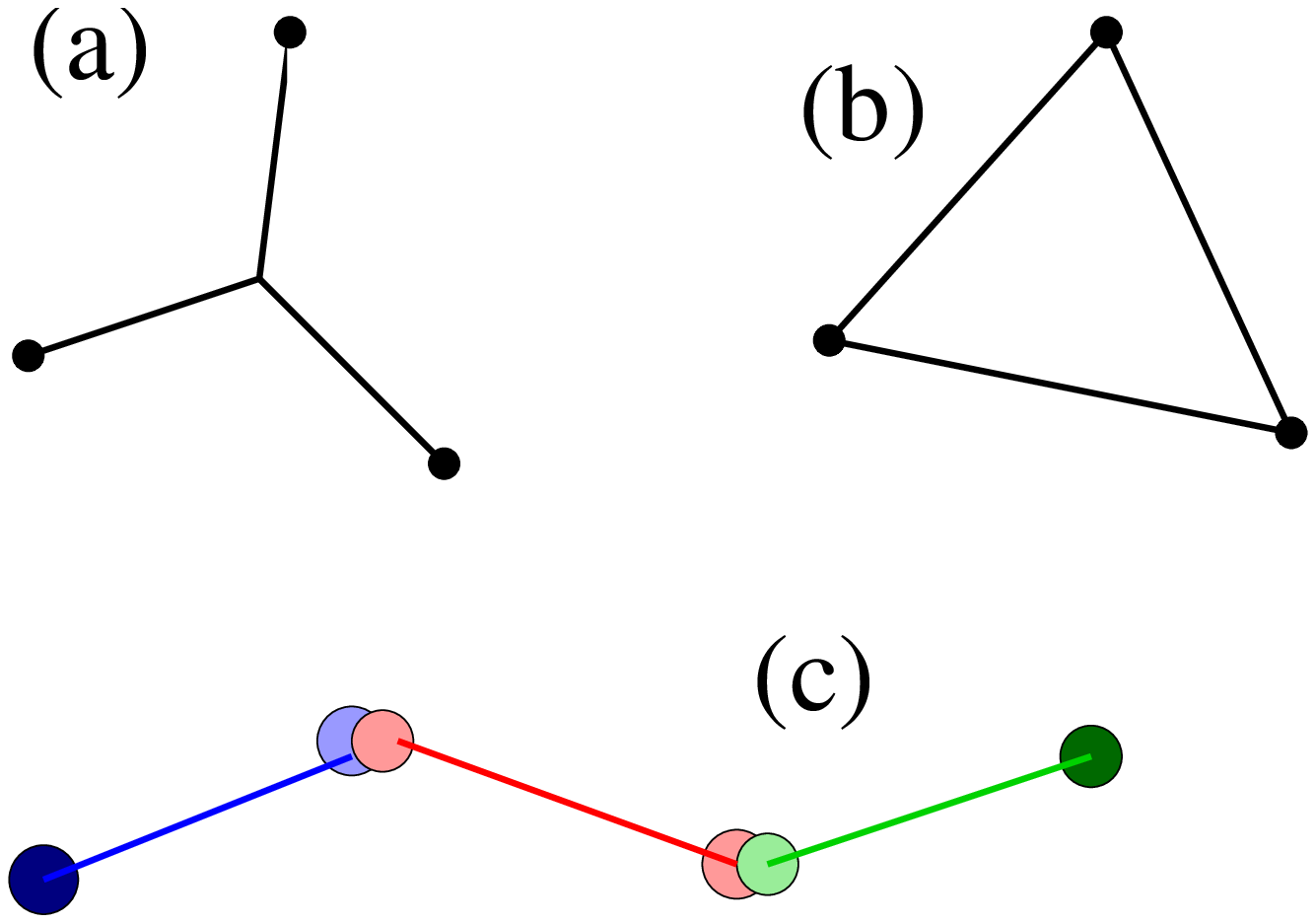}
\end{minipage}\hspace{1.5cm}
\begin{minipage}{8.cm}
\includegraphics[width=7cm]{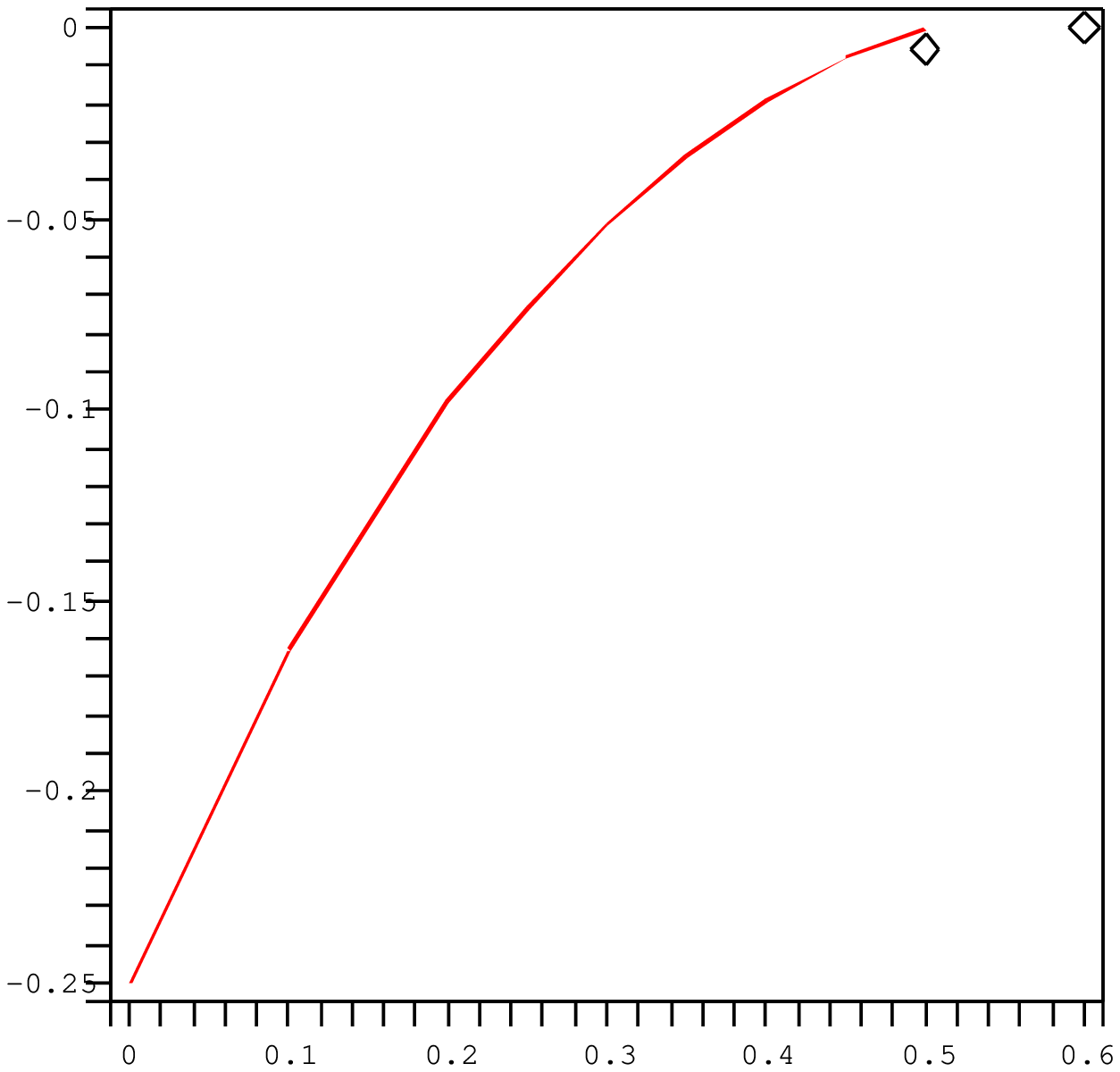}
\end{minipage}
 \caption{(left) The interaction in baryons for
``string-like'' interaction (a) versus the ``potential-like''
interaction (b). The double circles with different colors (online)
in (c) represent gluons, and it is an example of 4-chain $\bar q g
g q$. (right) Dependence of the binding energy on the Debye screening mass for
the simple exponential trial functions. The units are explained in
the text. Two diamond points indicate positions of the exact
solutions.
} \label{fig_baryon}
\end{figure}

   Applications of lattice-based binary potentials for static quarks
to multi-body problems meets a fundamental question: what part of
this interaction is (i) of a ``potential-type'' or (ii) of a
``string-type''. In the former case the potential energy of a
manybody state is the sum over $all$ pairs of particles, while in
the latter only some pre-defined partners, ``connected by a
 string'' are allowed to
interact.
The issue is well known and was discussed in literature for baryons
for decades. In Fig.\ref{fig_baryon}(left) we show how two pictures look
like, with strings in (a) ending at a string junction.

  The discussion above motivates us to consider interaction
to be string-like in this sense. An additional
 reason for that is the especially simple
multi-particle
states appear, namely the {\em polymeric chains}
made
of repeated gluons with $\bar q$ and $q$ at the ends, see
 Fig.\ref{fig_baryon}(c). (Recall that gluons have 2 color indices
and can be viewed as 2 different color charges connected to two
strings.)

For baryons, it turns out that the string-based picture (a) and
potential one (b) (with the Casimir factor 1/2 compared with meson
potential) give very close results, as is the case in vacuum
\cite{Baryon_vacuum}. Also, this is supported by recent lattice
study of free energy of static three quark
systems\cite{lattice_baryon} which found that the connected part
of qqq-singlet free energies above $T_c$ are decomposable into
three qq-triplet (diquark) free energies for all distances. We
will not  discuss more complicated possible structures, like
``polymerized'' baryons with extra gluons or a network of chains,
connected by color junctions.

 Although we would not calculate the
statistical and transport properties of sQGP as such in this
 work, we would comment on them at the end in the last section.

\section{The coordinates and the variational procedure}
We will denote particle coordinates by $\vec x_i,i=1,n$ where the
vector  means a 3-d space and the index is for the particle
number. Since the center of mass coordinate  does not appear in
the potential energy,  the corresponding momentum
 is conserved and can be put to zero. It is a
standard procedure to use some redefined coordinates  which
 reduce the n-body problem to the (n-1) one: those
coordinates would be denoted by  $\vec r_i,i=1,n$. Other notations
we will use are \be \vec\partial_i={\vec\partial \over\partial
x_i},  \vec D_i={\vec\partial \over\partial r_i}, \ee related by $
\vec\partial_i=\vec D_j M_{ji}$ where $M$ is the coordinate matrix
$r=Mx$ (namely $\vec r_j=M_{ji} \vec x_i$).

 For reference and further comparison, let us
give explicit form of the kinetic energy for (3-body) Jacobi
coordinates \be \label{jacobi} \vec r_1= \vec x_1-\vec x_2;  \vec
r_2= (-1/2)\vec x_1+(-1/2)\vec x_2+\vec x_3; \nonumber \\  \vec
r_3=(1/3)(\vec x_1+\vec x_2+\vec x_3) \ee
 \be \label{ke_jacobi} {1\over 2M}
(\vec\partial_i)^2|_{3Jacobi}= {1\over 2M}(2\vec D_1^2+(3/2)\vec
D_2^2+(1/3)\vec D_3^2) \ee
Note that
it is a diagonal form avoiding coordinate mixing.

However for the polymer problem we found that
more useful coordinates are  ``chain coordinates''
defined by the following set of $n$ coordinates
\be
 \vec r_1=\vec x_2-\vec x_1; ...  \vec r_{n-1}=\vec x_n-\vec x_{n-1}; \\ \nonumber
\vec r_n={1\over n}({ \vec x_1+...+ \vec  x_n})
\ee

The corresponding kinetic energies
%
%
 for N-chains is \be {1\over 2M}
(\vec\partial_i)^2|_{N-chain}= {1\over 2M}(2\vec D_1^2 + 2\vec
D_2^2+\cdot\cdot\cdot+2\vec D_{N-1}^2+ \\ \nonumber (1/N)\vec
D_N^2 -2 \sum_{i=1}^{N-2} \vec D_i \vec D_{i+1} ) \ee
Although it is not diagonal, it is very simple instead, with the
same ``reduced mass'' in each diagonal term\footnote{This of course
requires that all involved quasiparticles, the quark and the gluon ones,
 have the same mass, which is however approximately
fulfilled by available lattice data.}
.
 Furthermore, in ``string-like'' approach the potential energy is very simple,
just a
sum over all  ``bonds''  along the chain
\be E_{pot}=V(r_1)+...+V(r_{n-1})\ee
Note that we dropped
vector notation here: it means that only the lengths of
these coordinates matter.
Furthermore, all angular variables would not
be important for the ground s-wave states to be discussed.
If so, the  wave function  factorizes
\be \Psi(\vec r_1...\vec r_n)=\prod_{i=1}^{n-1} \psi(r_i)\ee
and as a result the average value of the
non-diagonal terms in the kinetic energy
would vanish
\be <\vec D_i \vec D_j>|_{i\neq j}=0\ee
Finally, as the diagonal terms have the same reduced mass as a
2-body problem, the problem obviously splits into (n-1)
Schroedinger equations.
 This completes the proof
that there is the same binding energy {\em per bond} (not per
particle) as  for mesonic states..
  The binding {\em per particle} of course grows, doubling the
binding in binary states as the length of the chain
grows\footnote{ We again remind the reader that it only happened
because in string-like approach one can ignore all the
interactions between  non-next-neighbors along the chains.}.

\section{Mesons and polymers in a variational approach}
  Since the 2-body problem and (as shown above) polymer chains
can be easily solved numerically for any potential,
and for relevant lattice-based potential it was already done
in \cite{SZ_bound},
the reader may be surprised why we discuss it here.
We however found  it  instructive to start with a simple example, for which
all calculations and integrals are simple and can be done analytically

Let the potential be just a screened Coulomb (or Yukawa) potential
\be V\, = \,-{\frac {\alpha\,{e^{-{\it M_D}\,r}}}{r}} \ee and  the
trial function be as simple as possible, namely an exponential
function \be {\it \psi }\, = \,{e^{-Ar}} \ee The average energy is
\be <H>&=&<(-1/2/M)*D^2-(1/r/M)*D+V(r)> \\ \nonumber &=&
1/2\,{\frac {{A}^{2}}{M}}
 -{4*\alpha*A^3\over (M_D+2A)^2}\ee
and it can be easily minimized in respect to parameter $A$.
The results (for $\alpha=1,M=1/2$) as a function of $M_D$
are plotted in Fig.\ref{fig_baryon}(right).

Note that the simple exponential trial function we use is exact
for a Coulomb problem ($M_D=0$ case), but of course is not so for
a screened potential. Although the energy may seem to be quite
close to exact ones, obtained from numerical solution of the
Schroedinger equation, true wave function is not particularly well
reproduced by it  as $M_D$ grows. In particular,  the curve in
Fig.\ref{fig_baryon}(right) crosses zero at its endpoint in a wrong
manner: in fact the curve must have a horizontal tangent at the
endpoint. Moreover, the critical value for the level disappearance
predicted by exponential trial function \be \label{endt} {\alpha
M\over M_D}|_{zero binding}=1 \ee
 is not at all accurate, as a comparison to exact behavior
(indicated by two points in the right upper corner of
Fig.\ref{fig_baryon})(right) changes the r.h.s. of (\ref{endt}) to a
smaller value, 5/6.

\begin{figure}
\begin{minipage}{8.cm}
\vskip -.3cm
\includegraphics[width=7.5cm]{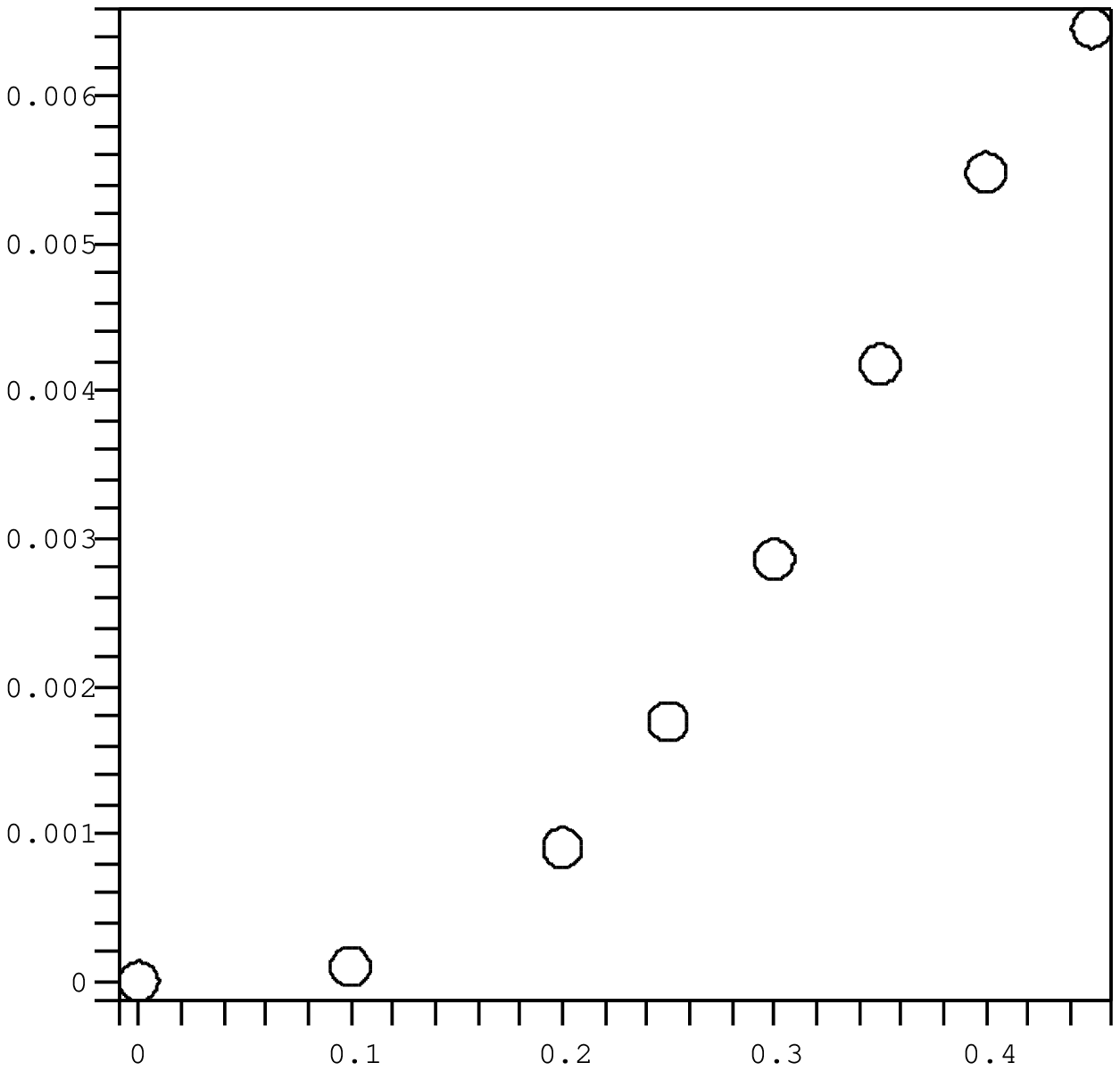}
\end{minipage}\begin{minipage}{8.cm}
\includegraphics[width=8.cm]{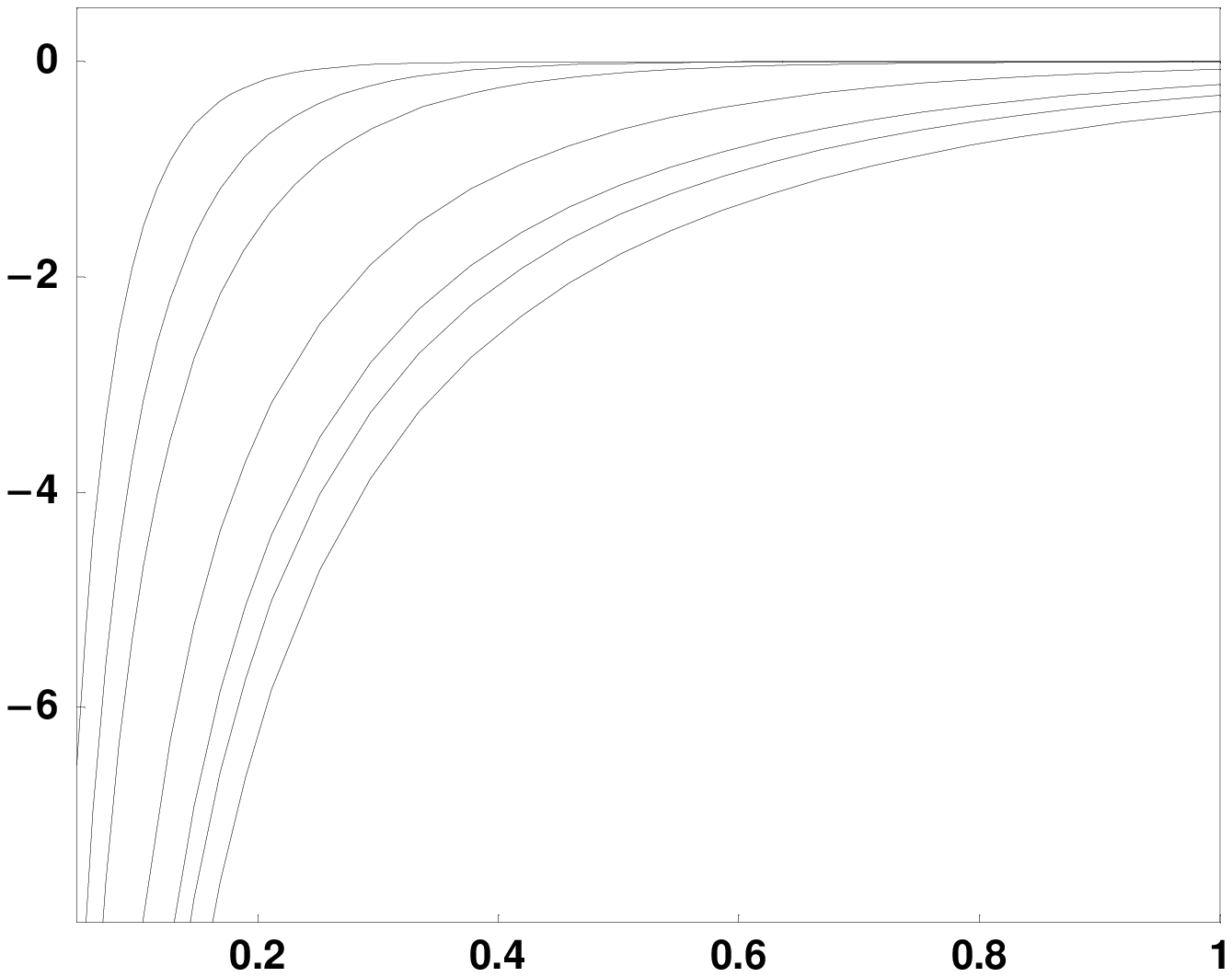}
 \end{minipage}
\caption{\label{fig_d}
(left) Dependence of the ``energy dispersion'' variable $d$ on the Debye
screening mass $M_D$, for the simple exponential trial functions.
(right)
The static potentials $V(T,r)$ (in unit $T_c$) as a function of
the distance r (in unit $1/T_c$). The values of temperature used
are $T=1,1.2,1.4,2,4,6,10 \, T_c$ for curves from right to left.}
 \end{figure}

One may ask whether it is possible to test the quality of the
trial wave function without a knowledge of the exact result (as
would be the case in multi-dimensional problems to be addressed in
subsequent section). A well known observable for that is the so
called ``energy dispersion'' variable, defined by \be
d=<H^2>-<H>^2 \ee If the trial wave function is an eigenvector,
d=0, otherwise it characterizes the quality of the approximation.
In Fig.\ref{fig_d} we show how it depends on $M_D$. As expected,
it is zero for pure Coulomb problem ($M_D=0$) but strongly grows
with $M_D$, indicating loss of quality of the approximation.
Nevertheless, we emphasize that while this energy dispersion is
quite sensitive to how close the trial function is to true
solution, the energy itself is much less sensitive to the details
of the wavefunction shape.

Now we turn to a realistic variational approach for the two-body
bonds in mesons and polymers, using a parameterized
temperature-dependent potential extracted from lattice data
\cite{SZ_bound} \be \label{potential}
V(T,r)=&-&(\frac{4}{3r}+\frac{8T}{3})\frac{e^{-2Tr}}{log(1+3T)}
\frac{4T}{r(1+3T)}\frac{e^{-2Tr}}{(log(1+3T))^2} \ee
Note that we have scaled all dimensional quantities with proper
powers of $T_c$, namely $T$ gives $T/T_c$, $r$ means $r T_c$, and
so on. We will keep these units throughout present paper. Plots of
the potential at different temperatures are shown in
Fig.\ref{fig_d}(right). At very short distance this potential
goes as a Coulomb (as expected from short-range one-gluon exchange
interaction) while at very large distance it is a screened Coulomb
potential which damps so fast as if it is almost vanishing.
The log term is simplified compared to the original form in
\cite{SZ_bound} where it was $log(1/r+3T)$, the parameterization
consistent with the asymptotic freedom at small $r$.

Note
that this is basically for color singlet $\bar{q} - q$ ( and for
color octet $g-g$ approximately), so we need to add appropriate
overall coefficients for other channels like diquark.

According to the features of the above potential, we employ a
trial wave function as following (with $\vec r$ the 2-body
relative coordinate) \be \label{trial} \phi (r) = e^{ -Ar-
\frac{C^2r^2}{(C^2r^2+1)} (B-A)r - \frac{1}{2}log(C^2r^2+1)} \ee It
has the asymptotic forms to be
\begin{eqnarray} \label{asymptotic}
 r\to 0 \,\, : \,\, \phi \to e^ {-A \, r} 
 r\to \infty \,\, : \,\, \phi \to \frac{e^{-B \, r}}{C \, r}
\end{eqnarray}
Here the parameter $C$ controls the interpolation between the
short distance Coulomb behavior and the long distance free
particle solution. All the three parameters have the same unit for
which we use $T_c$.

\begin{figure}
\includegraphics[width=7cm]{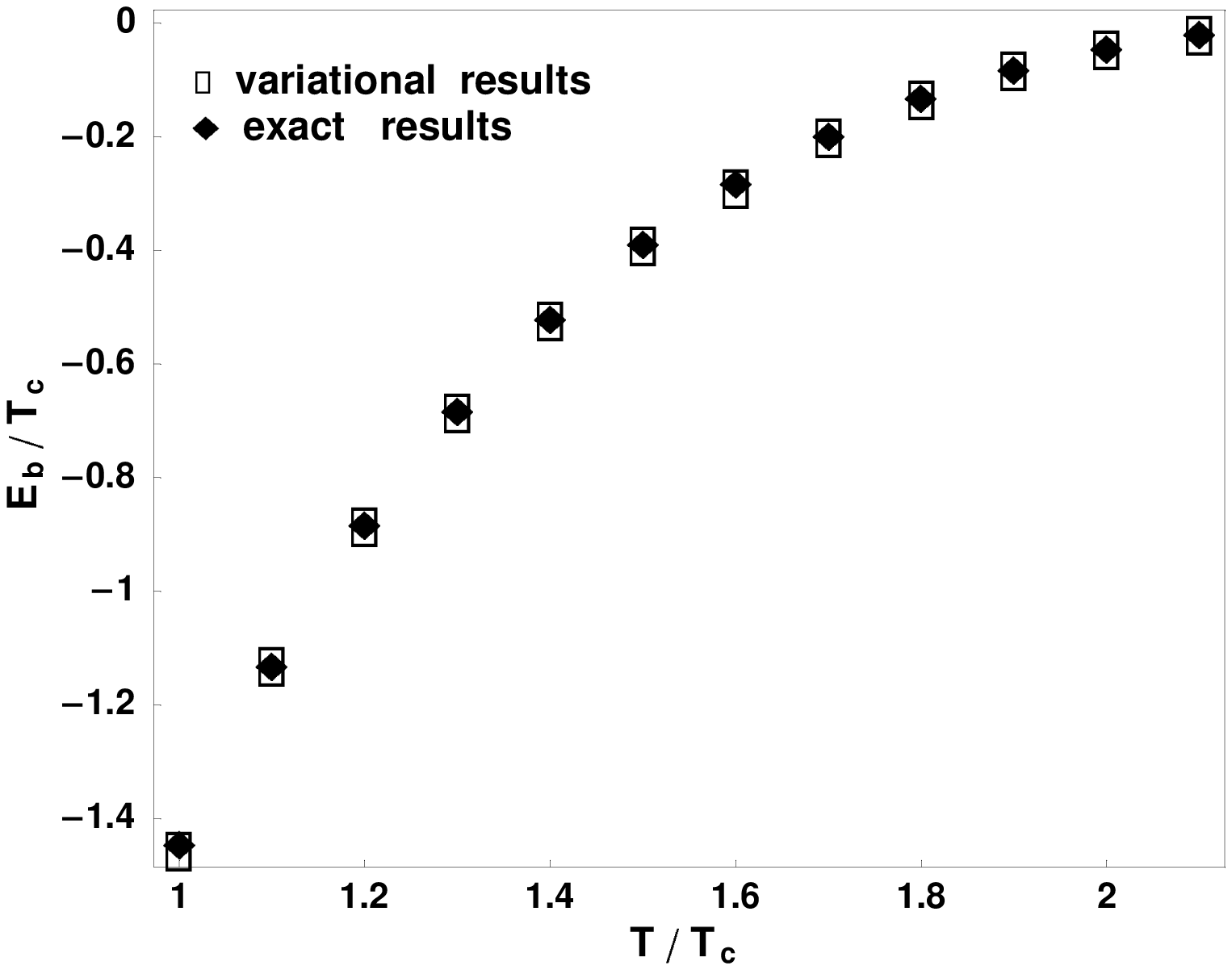}
\includegraphics[width=7cm]{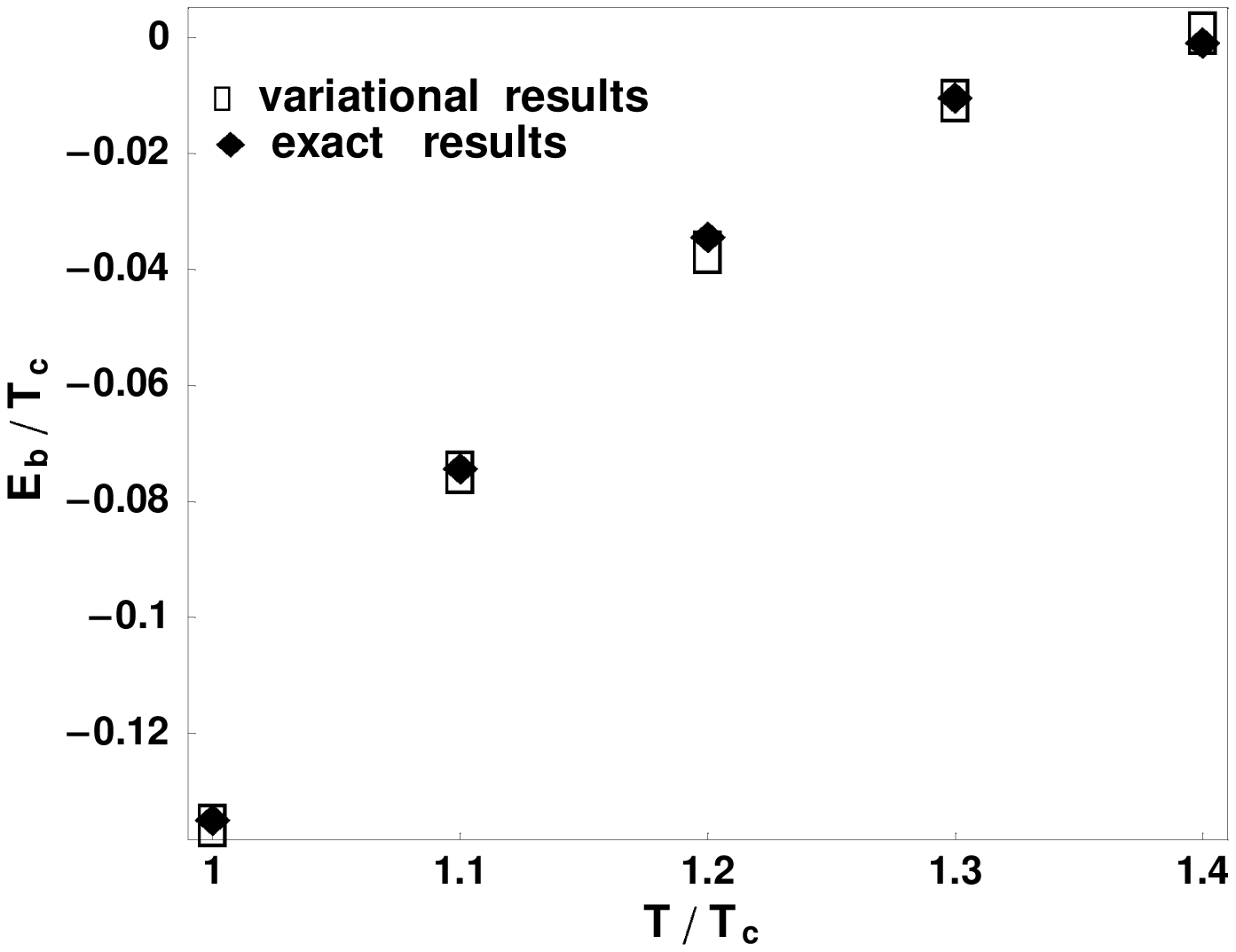}
 \caption{\label{fig_tmeson}
Dependence of the meson (left) and diquark (right) binding energy on the temperature. The
units are in $T_c$. }
 \end{figure}

With the (temperature-dependent) potential and trial function at
hand, we then find the binding energy at different temperature by
minimizing energy (kinetic plus potential energy) according to the
three variational parameters $A,B,C$. A non-relativistic form for
kinetic part is used, since the color-charged quasiparticles, both
quarks and gluons, are found by lattice simulations to be rather
heavy at temperatures not so high from $T_c$, for which we use the
same constant value $m^{*}=800MeV$ in our calculation for
simplicity. We employed the well-known Metropolis Monte
Carlo\cite{Metropolis} to evaluate the energy. The results are
shown in Fig.\ref{fig_tmeson}. We also plot the exact energies
obtained by numerically solving the Schroedinger equations for
comparison. As can be seen, the variational results coincide
perfectly with the exact results. The energy dispersions are of
order 0.01 (not vanishingly small), but as we emphasized before
and shown here, the energy itself could still be very accurately
evaluated. For the optimal values of wavefunction parameters, $A$
decreases from 3.125 to 2.25 and $B$ from 3.125 to 0.5 with
increasing temperature, and $C$ keeps between 0.5-1 to which the
energies are less sensitive.

The binding energy $E_b$ is greater than temperature for 1-1.1
$T_c$ and comparable with temperature up to 1.6 $T_c$($|E_b|/T$
about 0.26, and $e^{0.26}=1.3$). Naively following a $e^{-E_b/T}$
(at $T_c$ as high as 4.3) factor argument we may say the formation
of meson-like bonds in this T-region is quite favored. Now
recalling the proof in previous section, a polymer chain with N
elements will have a binding energy $(N-1)\* E_b$ ($E_b$ per
bond), and further more, the long chains have much more
statistical degeneracies (according to vast options arranging
intermediate gluons' quantum numbers), so we expect abundance of
polymer chains at temperatures just above $T_c$. And this, as we
will suggest, may dramatically contribute to jet quenching as well
as transport properties.

\section{Baryons at $T>Tc$}

After consideration of meson-like structures, it is of course
natural and interesting to address the question of possible
baryonic bound states above $T_c$. While lattice calculation found
mesonic bound states above $T_c$\cite{charmonium}, there is no
available information about baryon up till now. Also it is very
important to see the role of baryons in the deconfined phase of
QCD. For example, baryons carry conserved quantum numbers like
$B,S$ which may be combined to give useful experimental/lattice
signal\cite{LS_2} and contribute more to
 study of thermodynamics at non-zero chemical
potential\cite{Susceptibility}. So in this section we conduct a
variational approach for baryons in similar
way as was used in the past to study baryons in vacuum\cite{Baryon_vacuum}.\\

We consider baryons(anti-baryons) as closed 3-chains of
quarks(anti-quarks), which contains 3 pairs of
diquarks(anti-diquarks). Thus to study baryons, we first start
with diquarks at $T \ge T_c$. Diquarks in the deconfined phase of
QCD are of their own importance also. For the diquark channel, the
mutual interaction coupling should be one half the
quark-anti-quark channel, thus we adjust the potential
(\ref{potential}) by a coefficient $C/C_{\bar q q}=1/2$ to use it
for diquark bound states. We use the same trial function
(\ref{trial}) as in meson case, and then minimize energy according
to $A,B,C$. In Fig.\ref{fig_tmeson}(right) we plot the diquark binding
energy as a function of temperature. The exact numeric results are
also presented to compare and justify our variational approach.
Diquark
states are much more shallowly bound than meson and thus more easily melted.

For baryons, we construct the trial wavefunction on the basis of
the three diquark pairs. Since we're only interested in the ground
state, it is reasonable to use a totally symmetric s-wave spatial
configuration. The color wavefunction should still be the singlet
(then antisymmetric among the 3 quarks) to guarantee the
attractive interaction, but for spin and flavor part, the only
constraint is to be symmetric and there are a lot of ways to
arrange it (increasing statistical degeneracies). Particularly, we
want emphasize that very different from constructing baryon in
vacuum, now $s$ quark is more or less the same as $u,d$ quarks,
since its current mass, of the order of $T_c$ is much less than
quasiparticle mass and the current mass splitting is now
unimportant. So we write down the following \be \label{3bodytrial}
\psi(\vec x_1, \vec x_2, \vec x_3) = \phi(r_{12}) \phi(r_{23})
\phi(r_{31}) \ee
\begin{figure}
\includegraphics[width=7cm]{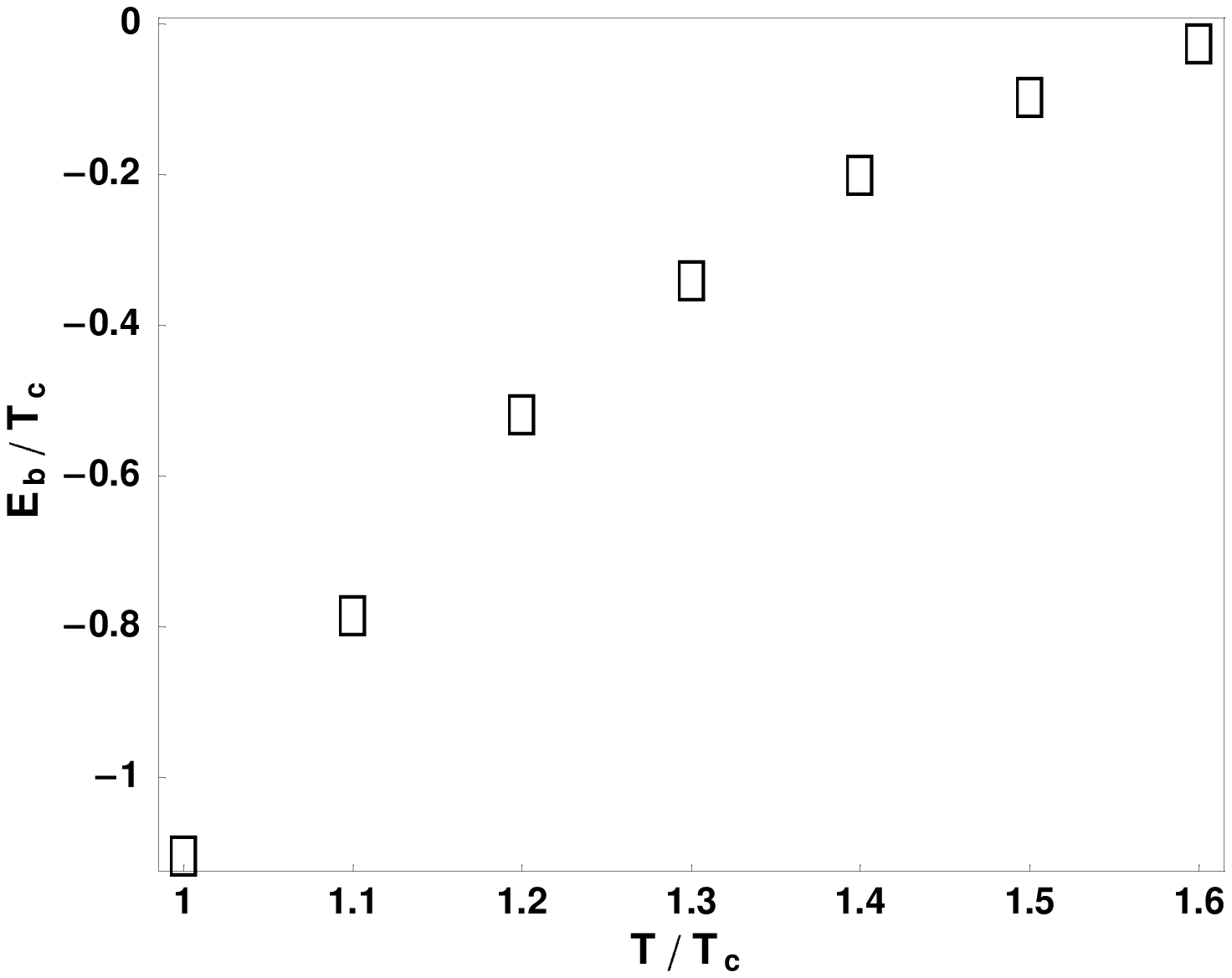}
\includegraphics[width=7cm]{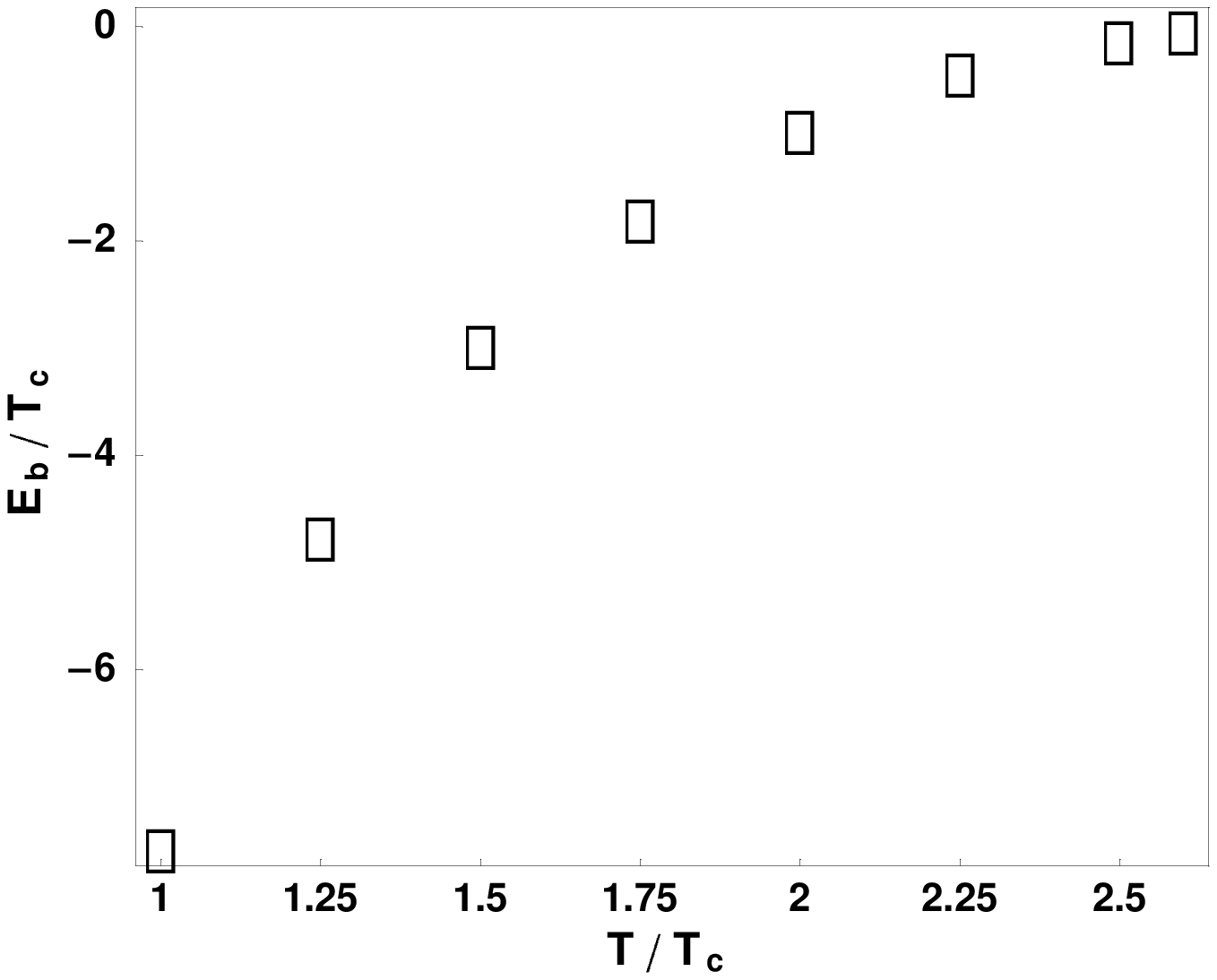}
 \caption{\label{fig_tbaryon}
Dependence of the baryon (left) and trigluon (right)
 binding energy on the temperature. The
units are in $T_c$.}
 \end{figure}

Here $\phi(r)$ is from (\ref{trial}) which has been used for meson
and diquark states, and $\vec r_{ij}=\vec x_i - \vec x_j$. Again
we minimize energy according to the variational parameters
$A,B,C$. Note now for this 3-body object, we have two kinetic
energy terms according to the first two in (\ref{ke_jacobi}) (one
term with reduced mass factor 1/2 and the other 2/3), while for
potential energy we need count for each pair of quarks, namely
three potential energy terms. Hence we expect that baryons are
more compact and deeply bound than diquarks both because of
heavier reduced mass and due to more potential energy, as can be
seen in Fig.\ref{fig_tbaryon}(left). The binding energy is slightly
greater than temperature at $T_c$ and comparable up to 1.3 $T_c$,
which means the baryons (and anti-baryons) should play some role
for temperatures not too high above $T_c$. As far as we know, this
point is noticed and demonstrated for the first time.

Having studied baryons, we go a step further and include the simplest
 closed 3-chain structure above $T_c$, namely $ggg$,  a
color-singlet channel in which the three gluons mutually interact
in similar way as three quarks in a baryon.  In a string picture,
       there is a string
between each pair, so the potential should be the same  as in a
meson (twice that for diquarks).  Again we use the 3-body trial
wavefunction (\ref{3bodytrial}) and minimize energy according to
$A,B,C$.  The results are shown in Fig.\ref{fig_tbaryon}(right).  It is
bound up to rather high temperature of about 2.6 $T_c$. The
binding energy at $T_c$ is as high as 7.64 $T_c$, and the size
shrinks to only about 0.3 fm. Since we only use static potentials,
without any relativistic corrections, we warn the reader that
close to $T_c$ the $ggg$ binding (the only one!) becomes too large
to be reliably evaluated inside the approximations
made\footnote{On
 top
of relativistic effects we ignored, the simplification of the potential
$log(1/r+3T) \to log(1+3T)$ used above also affects  binding close
to $T_c$. }.

\section{Summary and discussion}

\begin{figure}
\begin{center}
\includegraphics[width=10cm]{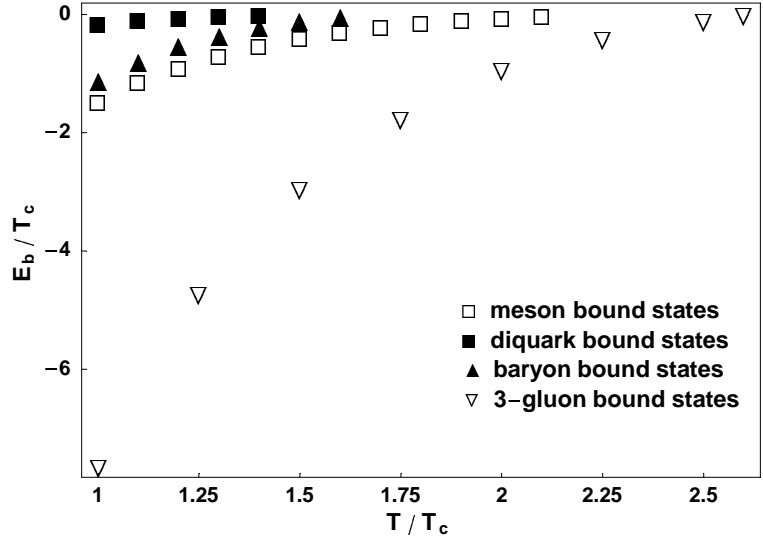}
\end{center}
 \caption{\label{fig_boundstates}
Dependence of various states' binding energy on the temperature.
The units are in $T_c$.}
 \end{figure}

To sum up, three {\em multibody} bound states have been studied
via variational approach: (i) ``polymer chains'' of the type $\bar
q g g ..g q$; (ii) baryons $(qqq)$; (iii) closed (3-)chains of
gluons $(ggg)$. For the chains (i) we have proved that they have
the same binding energy per bond as mesonic states and thus form
in the same temperature range as mesonic states. We have
established the binding energies and survival T-ranges for all
these three structures. All the results are summarized in
Fig.\ref{fig_boundstates} and Table.\ref{table_data}. We conclude
that between temperature region 1-1.5 $T_c$  the existence of all
these multibody bound states is not only possible but very robust.

\begin{table}[hptb]
\caption{\label{table_data} Summary of different bound states at
$T>T_c$ studied in this paper. The column $C/C_{\bar q q}$ gives
the relative potential strengths used in calculation, $E_b$ means
the binding energy, and $T_m$ refers to the melting temperatures
for different structures.}
\begin{tabular}{ccccc}
structure                       &  -body & $C/C_{\bar q q}$ & $E_b/T_c$ at $T_c$ & $T_m$  \\
 & & &  &  \\
 $\bar{q} q$                    &  2     & 1                           &  -1.45                  & 2.1 \\
 & & &  &  \\
 $\bar{q} g\cdot\cdot\cdot g q$\\(polymer chain) &  N     & 1                           &  -1.45$*$(N-1)            & 2.1  \\
 & & &  &  \\
 $ggg$\\(closed chain)            &  3     & 1                           &  -7.64            & 2.6  \\
 & & &  &  \\
 $qq$ / $\bar q \bar q$                         &  2     & 1/2                         &  -0.13                  & 1.4  \\
 & & &  &  \\
 $qqq$ / $\bar q \bar q \bar q$\\(closed chain)             &  3     & 1/2                         &  -1.10                  & 1.6
\end{tabular}
\end{table}

\begin{figure}
\begin{center}
\includegraphics[width=7cm]{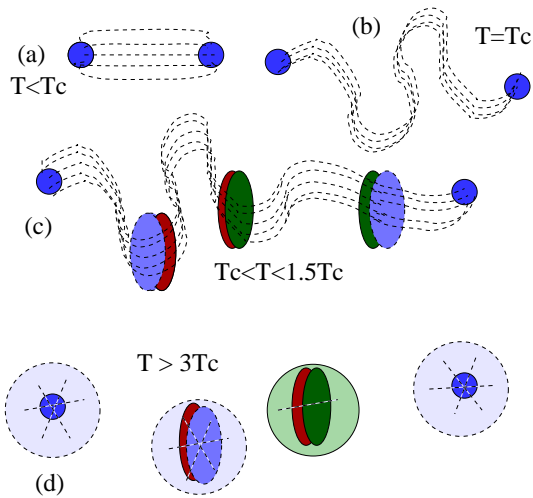}
\end{center}
 \caption{(color online)
A schematic picture of the distribution of color fields, at
different temperatures. Single color circles are quarks,
bi-colored ovals are gluons.
 } \label{fig_strings_polymers}
\end{figure}

Before we go forward with a general discussion, let us try to
summarize
the proposed scenario as a single picture, see Fig.\ref{fig_strings_polymers}.
From relatively short string-like configuration of color fields at low
$T$,
fig (a), one moves to longer strings (b) at the critical point
\cite{Polyakov:1978vu}. New is picture (c) which depicts ``polymeric
chains'' considered in this work, significant at $T=(1-1.5) T_c$.
Eventually, at high $T$, one goes into (d) with independent quark and
gluon
quasiparticles, neutralized by isotropic Debye clouds.

   We have not studied in this work neither more complicated
states, such as a hybrid of baryons and polymers
or a network of chains connected by color junctions, nor
have we attempted to evaluate the possible role of polymers/baryons
in thermodynamical and transport properties of sQGP (to be done elsewhere).

(In particular, we  show in \cite{LS_2}
that baryonic susceptibilities -- up to the 6-th derivatives over
chemical potential  -- studied recently
on the lattice by  the UK-Bielefeld collaboration \cite{Susceptibility}
provide good evidence in favor of existing of baryons near $T_c$, with
a mass rapidly increasing into the QGP phase.)

 Let us make here only few general comments on that.
The conbtribution of polymers into partition function can be
easily
evaluated via a geometric series: the resulting
 enhancement  factor, correcting  the
contribution of the $\bar q q$ mesonic states to those with any number
of intermediate gluons $\bar q g...g q$, is 
\be \label{eqn_poly} f_{polymers}={1 \over  
1-6 exp[(|\delta E|-M_g) / T] } \ee
where 6 is the color and spin degeneracy added by each link.
If one takes literally the bond binding we found, up to 1.4$T_c$, and
the effective gluon mass we used, one
finds that even at $T=T_c$ this factor only reaches about 1.2,
and rapidly decreases at higher $T$. 
The corresponding enhancement factor for baryons is a cube of that,
but baryons themselves are a small effect.
So, all this means that we think all multibody states we discussed
provide   just a few percent corrections into thermal properties
of sQGP \footnote{However if  some unincluded effects
would increase this bidning by about a factor of two, the zero in
 the denominator can be reached, forcing 
total ``polymerization'' of matter.}.

The reason we studied those states is to see
if they can be more
important for {\em transport properties}.
 It was first suggested by Shuryak and Zahed
 in \cite{SZ_rethinking}
that existence of marginal states must increase rescattering and
thus dramatically reduce viscosity (mean free paths), leading to a
collisional (hydrodynamical) regime of expansion. Since different
states get marginal at different $T$, one may hope this mechanism
to work at all $T$ up to about $2T_c$, the highest temperature at
RHIC.

 We will now argue that the situation can be different for
charm stopping or
 jet quenching. So far the estimates
 in the paper \cite{SZ_ionization} were only done
for jet energy losses due to ionization
of $binary$ mesonic states.
   The multibody bound clusters we studied above surely
should induce jet quenching even more drastically. This should be
especially true for
 {\em chain formation}, because it is known to be effective mechanism
of the momentum
distribution over larger volume of matter.
The chains, connected by junctions, can in principle even produce a
network.

 The effect of similar substructure on transport
 is well known in material sciences: in fact
an admixture of fibers is widely used to disperse forces applied to
the  system. The most famous examples are
  Kevlar fibers added to epoxy (or other plastic),
 now applied in wide range
of applications, from tires, boats etc. to such exotic ones as
 ``bullet-proof  vests'' and even   ``anti-mine boots''.

  Let us end up by speculating a bit about what would happen
if indeed only the ``well polymerized'' sQGP in the interval $T=(1-1.5)T_c$
 dominates the jet energy loss $dE/dx$. Since at RHIC
the initial $T$ is  about $2T_c$, jets would have
 smaller losses   till
the matter cools down and polymerizes properly.  Such a delay
is observable because jets can move out of matter during such latent time: this
 would  drastically affect the magnitude and especially angular distribution
of the jet quenching.

In fact the original idea of ``jet
tomography'' via jet quenching is  in serious trouble
for  quite a while, because
the most natural assumption --
the
energy loss  proportional to the matter density --
is in strong contradiction with
 the observed strong angular dependence of jet quenching \cite{v2_largept},
predicting  too weak azimuthal asymmetry for non-central
collisions. It was however recently pointed out by Pantuev
\cite{Pantuev:2005jt} that a better description of data can be
achieved $if$ the jet quenching at the highest RHIC energy is
switched on after some ``latent time'' of about 2.2 fm. This time
quite reasonably matches a cooling time from $T\approx 2 T_c$ to
$T\approx 1.5 T_c$ at RHIC.

{\bf Acknowledgments.\,\,} This work was partially supported by
the US-DOE grants DE-FG02-88ER40388 and DE-FG03-97ER4014.
We thank J.Casalderrey, V.Pantuev and K.Redlich for helpful discussions.


\end{document}